\documentclass[12pt]{article}
\usepackage{amssymb}
\usepackage{graphicx,color}
\usepackage{amsmath}
\usepackage{enumerate}
\usepackage{array}
\usepackage{xcolor}

\usepackage{epsfig}
\usepackage[english]{babel}
\usepackage{latexsym}

\usepackage{bm}
\usepackage{bm}
\def\eps{{\varepsilon}}

\newcommand{\fp}[1]{FP$ {\textrm{#1}}$}

\def\S{\mathcal{S}}

\def\eRM{{\mathrm e}}
\def\dRM{{\mathrm d}}

\def\mx{{\bm x}}
\def\mv{{\bm v}}

\def\mk{{\bm k}}

\def\mpp{{\bm p}}
\def\mr{{\bm r}}
\def\mJ{{\bm J}}
\def\mE{{\bm E}}
\def\mB{{\bm B}}

\def\eps{\varepsilon}
\def\boldnabla{{\bm \nabla}}

\allowdisplaybreaks
\allowbreak

\begin{document}

\title{Advection by Compressible Turbulent Flows: Renormalization Group Study of Vector and Tracer Admixture}

\author{N.~V.~Antonov$^{1}$, N.~M.~Gulitskiy$^{1}$, M.~M.~Kostenko$^{1}$, 
 T. Lu\v{c}ivjansk\'y$^{2}$}

\maketitle\mbox{ }
\\
 $^{1}$ Department of Physics, Saint Petersburg State University, \\ 7/9 Universitetskaya Naberezhnaya, Saint Petersburg 199034, Russia,\\ 
$^{2}$ Faculty of Sciences, P.J. \v{S}af\'arik
University, Moyzesova 16, 040 01 Ko\v{s}ice, Slovakia.

%----------------------------------------------------------------------------------------------------------------%
\begin{abstract}
Advection-diffusion problems of magnetic field and tracer field are analyzed using
the field theoretic perturbative renormalization group.
Both advected fields are considered to be passive, i.e., without any influence on the turbulent environment, and 
advecting velocity field is generated by compressible version of stochastic Navier-Stokes
 equation. The model is considered in the vicinity
of space dimension $d = 4$ and is a continuation of previous work [N.V. Antonov et al., Phys. Rev. E {\bf 95}, 033120 (2017)]. 
The perturbation theory near the special dimension $d=4$ is constructed within a double
expansion scheme in $y$ (which describes scaling behavior of the
random force that enters a stochastic equation for the velocity field) and $\eps=4-d$. We show that 
up to one-loop approximation both types of advected fields exhibit similar universal scaling behavior. In particular,
we demonstrate this statement on the inertial range asymptotic behavior of the correlation functions of advected fields.
The critical dimensions of tensor composite operators are calculated in the leading order of $(y,\eps)$ expansion.
 
%---------------------------------------------------------------------------------------------------%
\subsection*{Keywords}
{fully developed turbulence, magnetohydrodynamics, advection-diffusion problem, field-theoretic
renormalization group, anomalous scaling}

\end{abstract}

%----------------------------------------------------------------------------------------------------------------%
{\section{Introduction} 
\label{sec:intro}}

%===========================================================================================
%		General Intro
%===========================================================================================
Many natural phenomena involve broad range of spatial or time scales. For instance, 
continuous phase transitions, diffusion-driven systems, population dynamics or turbulent flows provide
famous examples~\cite{Vasiliev,Tauber,Zinn}. 
Both from theoretical and practical point of view turbulence plays a distinguished role.
Due to a relatively low value of air viscosity~\cite{Frisch,Monin,Davidson}, turbulent flows
are much more easily generated than is commonly believed.
Despite a vast amount of efforts that
has been put into, 
the fundamental understanding of turbulence remains unsolved and it is widely
regarded as a last unsolved classical problem. 
Arguably, turbulence exhibits many interesting features, such as scaling behavior, prominent intermittency,
coherent structures and others~\cite{Frisch,Davidson,Falkovich2001}.
A distinctive quantitative aspect is known as anomalous scaling, i.e., singular power-law dependence of outer scale $L$ of some statistical quantities 
in the inertial-convective range~\cite{Frisch,Monin}. 
Its proper investigation requires a lot of thorough analysis. 
The general aim of theory is to predict possible macroscopic behavior of a turbulent fluid
and to give a quantitative prediction about characteristic quantities (correlation and structure functions).
%===========================================================================================
%		astrophysical context
%=========================================================================================

In astrophysics turbulence plays probably even more
important role than in terrestrial events~\cite{biskamp,Shore,Priest}. Being a 
mechanism for an explanation of many
effects: magnetic dynamo in interior of planets, convective processes in stars, outbreaks of prominences on
Sun's surface, galaxy formations and others~\cite{pouquet,tu,balbus,chabrier,elmegreen, feder}, it is clear
that mutual interplay between turbulent flow and additional advected field is quite common in nature.
Well-known  model for a theoretical description of magnetohydrodynamic (MHD) is so-called Kazantsev-Kraichnan kinematic model~\cite{ant06}. 
Its basic premise is to assume that a magnetic vector field (later in this article
referred just as a vector field) is passively advected by
turbulent velocity field with no backward influence to the velocity field (for
a general introduction to magnetohydrodynamic see, e.g.,~\cite{moffatt}). 
Thus, the Kazantzev-Kraichnan model can be viewed as a simplified version to the full MHD
problem, in which the Lorentz force is neglected.
There are many studies~\cite{Four82,AVH84} devoted to this problem, mainly
because it provides a mechanism for a generation of turbulent dynamo~\cite{biskamp,moffatt}.
The main point of criticism on Kazantsev-Kraichnan model is an assumption of the 
velocity field, which according to this model is simply given by a Gaussian random variable.
More appropriate approach would  consider velocity field to be generated by some dynamical mechanism. 
 
As a rule, in astrophysical context we are dealing with a compressible fluid
rather than incompressible~\cite{Shore}. 
Here, we therefore employ a
compressible version of Navier-Stokes equation (NS) for a generation
of velocity fluctuations~\cite{LL,Sagaut} and study its effect on an advection of magnetic field.
Such (compressible) MHD models witnessed a considerable
scientific activity in recent years~\cite{Kim05,Carbone09,Sahra09,Eyink10,Galtier11,Banerjee13,Banerjee16,Hadid17}.

This work is motivated by
the previous studies~\cite{Iroshnikov,AG12,AG12-2,AG12-3,J13,Uni3,Uni3-2,VectorN,VectorN-2,VectorN-3,VectorN-4} of
the incompressible case.  
%===========================================================================================
%		tracer admixture
%=========================================================================================
Besides advection problem of vector quantity, we consider in this paper also
a problem of passive scalar. In particular, we have in mind tracer field
advected by the aforementioned compressible version of stochastic Navier-Stokes eqaution. 
The reason is that as concrete calculations shows, results for vector and tracer case
share similarities. As we will see, concrete expressions for universal quantities will be the same up to some factor. 

%===========================================================================================
%		RG approach
%=========================================================================================
The investigation of such behavior as anomalous scaling requires a lot of thorough analysis to be carried out.
The phenomenon manifests itself
in a singular intermittent behavior of some statistical quantities (correlation 
and response functions, structure functions, etc.) in the
inertial-convective range in the fully developed turbulence regime~\cite{Frisch,Davidson,Falkovich2001}.
As has been mentioned previously, turbulent flows are accompanied by a lack of typical scale.
This shares a somewhat formal similarity with a physics related to critical phenomena. 
A very useful and computationally effective approach to the problems with many interacting degrees of 
freedom on different scales is the field-theoretic 
renormalization group (RG) approach which can be subsequently accompanied by the operator product
expansion (OPE); see the monographs~\cite{Vasiliev,Tauber,turbo}. 

It is a difficult problem to investigate both the Navier-Stokes equation for a compressible 
fluid and passive advection problems by this velocity ensemble. The first relevant discussion and analysis 
of passive advection emerged a few 
decades ago for the Kraichnan's velocity ensemble which modelled advection of impurity by incompressible fluid~\cite{Kraich1,GK,RG}. Further 
studies developed its more realistic generalizations~\cite{AG12,AG12-2,AG12-3,J13,Uni3,Uni3-2,VectorN,VectorN-2,VectorN-3,VectorN-4,Ant99,JJ15,V96-mod,amodel,JJM18,HHL} 
and, in particular, to the 
compressible case~\cite{NSpass,Ant04,AGM,JJR16,JJM17,VM,tracer2,tracer3,AdzAnt98,ANU97,AK14,AK15,AK2,LM,St,MTY}.

As we will see, studied models of compressible fluid reveal intriguing behavior
near the specific space dimension $d=4$. Usually, $d$ plays a passive role in advection problems,
but sometimes may affect the RG procedure: consideration of compressible fluid near $d=4$ is very close to analysis of
incompressible fluid near special space dimension  $d=2$. In this case an  additional divergence appears
in the 1-irreducible Green function $\left\langle v'v'\right\rangle_{\text{1-ir}}$,
see~\cite{HN96,AHKV03,AHH10}. This feature allows us to employ a double expansion scheme, in
which the formal expansion 
parameters are $y$, which describes the scaling behavior of a random force, and $\eps=4-d$,
i.e., a deviation from the space dimension $d=4$.

%===========================================================================================
%		Organization of the paper
%===========================================================================================
The paper is organized as follows. First, we begin with a description of compressible fluid dynamics in  Section~\ref{sec:NSmodel}.
Then, in Section~\ref{sec:advection} we proceed to a description of advection problem of passive tracer quantity and magnetic (vector) field, respectively.
In Section~\ref{sec:field_model} we reformulate studied models into a field-theoretic formalism, which
is subsequently analyzed in Section~\ref{sec:RG}.
Discussion of the fixed points' structure and related  scaling regimes is presented in Section~\ref{sec:points}.
In Section~\ref{sec:CF} the renormalization of a certain composite fields is considered and anomalous exponents are calculated.
In Section~\ref{sec:OPE} OPE is applied to the various correlation functions. The concluding Section~\ref{sec:end} is
devoted to the brief discussion.   

%-----------------------------------------------------------------------------------------------------------%
{\section{Navier-Stokes for compressible fluid} 
\label{sec:NSmodel}}

A quantitative parameter that describes intensity 
of turbulent motion is so-called 
Reynolds number $\mathrm{Re}$ {which
represents} a ratio between inertial and dissipative forces~\cite{Frisch,Davidson,LL}. For high enough values of 
$\mathrm{Re} \gg 1$ inertial interval emerges in which just transfer of kinetic energy from outer $L$ (input) to microscopic $l$ (dissipative) scales
take place. One of the microscopic models used for a description of fully developed turbulence in inertial interval is based on a stochastic version
of Navier-Stokes equation~\cite{Vasiliev,turbo}.
According to it the
 dynamics of a compressible fluid is governed by the following equation~\cite{LL,ANU97}
\begin{equation}
  \rho\nabla_{t} v_{i} = \nu_{0} [\delta_{ik}\partial^{2}-
  \partial_{i}\partial_{k}] v_{k}
  + \mu_0 \partial_{i}\partial_{k} v_{k} - \partial_{i} p + f^v_{i},
  \label{eq:NS}
\end{equation}
where the operator $\nabla_t$ denotes Lagrangian convective derivative $\nabla_{t} = \partial_{t} + v_{k} \partial_{k}$, $\rho=\rho(t,\mx)$ is a fluid 
density field, $v_i=v_i(t,\mx)$ is the velocity field, $p=p(t,\mx)$ is the pressure field, and $f^v_i$ is the external force, $\partial_{t} =
\partial /\partial t$ is a time derivative, $\partial_{i} = \partial /\partial x_{i}$ is 
$i$-th component gradient, and $\partial^{2} =\partial_{i}\partial_{i}$ is the Laplace operator. 
Two parameters $\nu_{0}$ and $\mu_{0}$ in Eq.~(\ref{eq:NS}) are two viscosity coefficients~\cite{LL}.
In this work we use a shorthand notation in which summations over repeated vector indices (Einstein summation convention) are always implied. 
 In subsequent sections we employ RG method, in which it is necessary to distinguish between unrenormalized (bare) and 
renormalized parameters. The former we denote by a subscript ``0.''
 
Let us note two important remarks regarding the physical interpretation of Eq.~(\ref{eq:NS}).
First, velocity field $v_i$  should 
be regarded as  a fluctuating part of the total velocity field. In other words, it is implicitly assumed that the mean (regular) part of the velocity
field has been already subtracted~\cite{Frisch,Davidson}. This point of view reflects philosophy behind the theory of critical phenomena, where
order parameter fluctuates around certain mean value as well~\cite{Vasiliev,papo}. Second, the random force $f_i^v$ accounts for two underlying physical
processes: a) continuous input of energy, which is needed in order to compensate losses of energy due to viscous terms in
Eq.~(\ref{eq:NS}), and b) interactions between fluctuating part of the velocity and the regular mean flow~\cite{Davidson,turbo}. 
 
To conclude the theoretical setup of velocity field, Eq.~(\ref{eq:NS}) 
has to be supplemented by two equations: a continuity equation 
\begin{equation}
\partial_{t} \rho  + \partial_{i} (\rho v_{i})   = 0
\label{eq:CE}
\end{equation}
and an additional relation coming from thermodynamic considerations 
\begin{equation}
\delta p = c_0^2 \delta\rho.
\label{eq:SE}
\end{equation}
Here,  $\delta p = p - \overline{p}$ and $\delta \rho = \rho-\overline{\rho}$ give deviations from the
equilibrium (mean) values of pressure field $\overline{p}$
and density field $\overline{\rho}$, a parameter $c_0$ is the adiabatic speed of sound.
 
Viscous terms in Eq.~(\ref{eq:NS}) proportional to parameters $\nu_{0}$ and $\mu_{0}$ characterize dissipative processes, which
are predominantly relevant at small spatial scales. Without a continuous input of energy 
turbulent processes necessarily fade away and {the flow eventually becomes} regular.  There are several possibilities for theoretical
description of energy input~\cite{turbo,FNS}. It is advantageous to define properties of the random force $f_i^v$ in frequency-momentum representation
\begin{equation}
\langle f_i^v(t,{\mx}) f_j^v(t',{\mx}') = \frac{\delta(t-t')}{(2\pi)^d} 
\int\displaylimits_{k>m} \dRM^d k \mbox{ }
D^v_{ij}({\mk})
\eRM^{i{\mk}\cdot({\mx-\mx'})},
\label{eq:ff}
\end{equation}
where the delta function ensures Galilean invariance of the model~\cite{turbo}. The integral
is  infrared~(IR) regularized with a parameter $m\sim L_v^{-1}$, where 
$L_v$ denotes outer scale, i.e., scale of the largest turbulent eddies~\cite{turbo,JETP}. 
Parameter $d$ denotes dimensionality of space. In what follows $d$ will be considered
as a continuous parameter in a  dimensional regularization, therefore we write it explicitly and do not immediately
insert its realistic three-dimensional value.
The kernel function ${D}^v_{ij}({\bm k})$ reads
\begin{equation}  
D_{ij}^v({\bm k})=g_{10} \nu_0^3 k^{4-d-y} \biggl\{
P_{ij}({\bm k}) + \alpha Q_{ij}({\bm k})
\biggl\} + g_{20} \nu_0^3 \delta_{ij}.
\label{eq:correl2}
\end{equation}
The non-local term proportional to the charge $g_{10}$ is chosen in a power law form 
that facilitates application of RG method. Dimensionless parameter $\alpha$ measures
an  intensity with which energy flows into a system via longitudinal modes~\cite{feder}.
Scaling exponent $y$ measures a deviation from a logarithmic behavior 
achieved for $y=0$. Moreover, it is possible to obtain 
a perturbative expansion in formally small $y$~\cite{Vasiliev,JETP}.
Stochastic theory of turbulence is mainly interested in the limiting case
$y\rightarrow 4$ that corresponds to an idealized input of energy from infinite spatial scales~\cite{Frisch,turbo}.
The transverse and longitudional projection operators $P_{ij}$ and $Q_{ij}$
in the momentum space read
\begin{equation}
P_{ij} (\mk) = \delta_{ij} - {k_i k_j}/{k^2}, \quad Q_{ij}(\mk) = {k_i k_j}/{k^2};
\label{eq:project}
\end{equation}
$k=|\mk|$ is the 
wave number. As we will see in Section~\ref{sec:RG} 
the presence of local term in~(\ref{eq:correl2}) is imposed by the renormalizability
considerations~\cite{AGKL17_pre,AGKL16,AGKL17_epj,AGKL17_epj2}. 
Effectively, presence of two charges leads to a double expansion scheme in $(y,\eps)$, where
$y$ has been introduced in Eq.~(\ref{eq:correl2}), and $\eps=4-d$, i.e., $\eps$
gives a deviation from the space dimension $d=4$~\cite{HHL}. 

%-----------------------------------------------------------------------------------------------------------%
{\section{Stochastic formulation of advection models} 
\label{sec:advection}}

In this section we briefly describe differential equations that govern advection of impurity fields by some velocity flow:
time evolution
of magnetic field in so-called Kazantsev-Kraichnan model and dynamics of simple tracer admixture.

%--------------------------------------------------------------------------------------------------------
%		INTRO OF MAGNETIC FIELD
%--------------------------------------------------------------------------------------------------------
The inclusion of magnetic field in Kazantsev-Kraichnan model follows a simple
physical considerations called magnetohydrodynamic limit~\cite{biskamp,Shore}. We assume that the medium is completely neutral at macroscopic scale 
and that free path of the particles is much smaller than Debye length. Therefore, we may 
neglect the displacement current, which 
is responsible for bulk motion of the ions and electrons, and describe our system in the bulk variables of density, pressure, and mean velocity fields only.
From the technical point of view and RG principles displacement current is IR irrelevant and, therefore, we do not need to preserve it in our model.
Taking into account Faraday's law $\partial_t\mB = -\boldnabla \times\mE$ together with
a generalized Ohm's law for a conducting fluid in motion $ {\mJ} = \sigma({\mE} + \mv\times{\mB})$ one gets
advection-diffusion equation $\partial_t \mB -\boldnabla\times(\mv\times\mB)=\kappa_0\boldnabla^2\mB$.
In a similar philosophy to Sec.~\ref{sec:NSmodel} stochastic version then 
takes the following form
\begin{equation}
\partial_t \theta_i + \partial_k(v_k\theta_i - v_i \theta_k) = \kappa_0 \partial^2 \theta_i + f^\theta_i,
\label{eq:eom_mag}
\end{equation}
where $\theta_i$ is a fluctuating component of total magnetic field,
$\kappa_0$ is the magnetic diffusion, and we have added stochastic term $f^\theta_i$ on the right hand side being the random component of the curl of current and stemming from intrinsic stochasticity of the magnetic field~\cite{turbo}.
Detailed exposition of the MHD equation can be found in the literature~\cite{biskamp,Shore,moffatt}. 
Let us note that in stochastic approach to MHD Eq.~(\ref{eq:eom_mag}) should be understood as
an equation for the fluctuating part $\theta_i=\theta_i(t,\mx)$ of the total magnetic 
field $B_i$, i.e., $B_i=B^0(n_i+\theta_i)$ with $B^0_i=B^0n_i$ and ${\bf n}$ being a constant background field  and a constant unit vector, respectively~\cite{Four82,AVH84,ALM00,Zhou2}.

Random force $f_i^\theta$ in Eq.~(\ref{eq:eom_mag}) is assumed to be a Gaussian random variable
with zero mean and given covariance,
\begin{equation}
\langle f_i^\theta(t,\mx) f_j^\theta (t',\mx') \rangle = \delta(t-t')\, C_{ij}({\mr}/L_\theta), \quad
{\mr}= {\mx} - {\mx}',
\label{eq:noise1}
\end{equation}
where $C_{ij}({\bf r}/L_\theta)$ is a function, whose precise functional form is unimportant. 
It has a finite limit at $({\mr}/L_\theta)\to 0$ and it rapidly decays for
$({\mr}/L_\theta)\to\infty$.
 Magnetic
field $\theta_i$ is divergence-free, which yields an equality between terms
$\partial_{k}(v_{i}\theta_k)$ and $(\theta_k \partial_{k})v_i $.

In more realistic scenarios there should be an additional Lorentz term in Eq.~(\ref{eq:NS}), which
corresponds to the active advection of magnetic field. This would require presence of the
Lorentz term $\mv\times\mB \sim {\bm J} \sim (\boldnabla\times \mB)\times \mB$, 
which would affect dynamics of velocity field and the resulting model would contain two interconnected
stochastic differential equations. However, this is beyond the scope of the present paper.
Moreover, it was found that in some special cases the only IR attractive fixed point in full model corresponds to passive (not active) advection of impurity fields~\cite{Four82,AVH84,NB98}.
 
%--------------------------------------------------------------------------------------------------------
%		INTRO OF TRACER FIELD
%--------------------------------------------------------------------------------------------------------
Thus, model~(\ref{eq:eom_mag}) corresponds
to a model of passive advection of magnetic field, which we later refer to as a vector model.
Related problem can be considered for a case
of scalar quantity $\theta=\theta(t,\mx)$ which represents 
the density of some pollutant, temperature field, concentration, etc. 
There are two permissible kinds of passive scalar fields in nature:
the density field (density of some pollutant) and the tracer field which describes the temperature or entropy~\cite{LL}.
The advection of a density field is governed by equation
\begin{equation}
\partial _t\theta+ \partial_{i}(v_{i}\theta)=\kappa _0 \partial^{2} \theta+f^\theta,
\end{equation} 
whereas advection of a tracer field is governed by equation
\begin{equation}
\partial _t\theta+ (v_{i}\partial_{i})\theta=\kappa _0\partial^{2} \theta+f^\theta;
\label{eq:tracer1}
\end{equation}
here in both equations $\kappa_0$ is the corresponding molecular diffusivity coefficient and $f= f(t,\mx)$ 
is again a Gaussian random variable with zero mean and given covariance,
\begin{equation}
\langle f(t,\mx)f(t',\mx') \rangle = \delta(t-t')\, C({\mr}/L), \quad
{\mr}= {\mx} - {\mx}'.
\label{eq:noise2}
\end{equation}
The function $C$ in Eq.~\eqref{eq:noise2} meets same criteria as function $C_{ij}$ in Eq.~(\ref{eq:noise1}).
For the incompressible fluid the density and tracer advection problems
are identical since transversality condition $\partial_i v_i = 0$
makes expressions $\partial_{i}(v_{i}\theta)$ and $(v_{i}\partial_{i})\theta$ equal, but for the compressible flows differences might appear~\cite{Ant00}.
The case of density advection was considered earlier in~\cite{AGKL17_pre,AGKL17_epj}; the case of tracer field is considered here together with vector model. 

%-----------------------------------------------------------------------------------------------------------%
{\section{Field-theoretic formulation} 
\label{sec:field_model}}

The main aim of this study is to investigate the scaling behavior of various statistical quantities
 (Green functions) of the theory near the special space dimension $d=4$.
In statistical physics we are interested in the macroscopic large-scale behavior
that corresponds to the IR range. Our main theoretical tool is the renormalization group theory, 
 which allows us to identify 
scaling regimes and analyse certain composite operators. 
An important difference of the present study with the traditional approaches 
is a special role of the space dimension~$d=4$.

Fortunately, despite the obvious differences between the stochastic formulations for vector and tracer fields 
[compare Eqs.~(\ref{eq:eom_mag}) and~(\ref{eq:tracer1})], there exist some similarities which allows us to perform their 
RG analysis at once.
In order to derive renormalizable field theory, the  stochastic equation~(\ref{eq:NS}) 
has to be 
divided by $\rho$, and fluctuations in viscous terms have to be neglected~\cite{VN96}.
Further, by using the expressions~\eqref{eq:CE} and~\eqref{eq:SE}
the problem can be recast in the form of two coupled equations:
\begin{align}
\nabla_{t} v_{i} & = 
\nu_{0} [\delta_{ik}\partial^{2}-\partial_{i}\partial_{k}]
v_{k}\! +\! \mu_0 \partial_{i}\partial_{k} v_{k} -\!
\partial_{i} \phi\! +\! f_{i},
\label{eq:ANU} \\
\nabla_{t} \phi & =  -c_{0}^{2} \partial_{i}v_{i}.
\label{eq:ANU1}
\end{align}
Here, a new field $\phi=\phi(t,\mx)$ has been introduced and it is related to the density fluctuations via the 
relation $\phi = c_0^2 \ln (\rho/\overline{\rho})$~\cite{AGKL17_pre,VN96}, and 
$f_{i}=f_{i}(t,\mx)$ is the external force normalized per unit mass.

According to the general theorem~\cite{Vasiliev,Zinn},  stochastic problems summarized by Eqs.~(\ref{eq:eom_mag}), (\ref{eq:ANU}), (\ref{eq:ANU1}) and 
Eqs.~(\ref{eq:tracer1}),~(\ref{eq:ANU}),~(\ref{eq:ANU1}), respectively,
are equivalent to the field theoretic
models with a doubled set of fields and 
certain De Dominicis-Janssen action functional~\cite{Janssen76,deDom76,Janssen79}. In the case of Kazantsev-Kraichnan model 
it is given by a sum of two terms 
\begin{align}
\S_\text{1} & = \S_\text{vel} + \S_\text{mag}, 
\label{eq:full_action1} 
\end{align}
where $\S_\text{vel}$ describes a velocity part
\begin{align}
\S_\text{vel} & = \frac{v_i' {D}_{ij}^v v_j'}{2} 
+v_i' \biggl[
-\nabla_t v_i + \nu_0(\delta_{ij}\partial^2 - \partial_i \partial_j)v_j
+u_0 \nu_0 \partial_i \partial_j v_j - \partial_i \phi
\biggl] \nonumber \\  &
+\phi'[-\nabla_t \phi  + v_0 \nu_0 \partial^2 \phi - c_0^2 (\partial_i v_i)]
\label{eq:vel_action} 
\end{align}  
with $ {D}^v_{ij}$ being the correlation function~(\ref{eq:correl2}). Note that we have introduced 
a new dimensionless parameter
$u_0=\mu_0/\nu_0>0$ and a new term
$v_0 \nu_{0} \phi' \partial^{2}\phi$ with another positive
dimensionless parameter $v_0$, which is needed  to ensure 
multiplicative renormalizability. Also we employ a condensed notation, in which integrals over the spatial variable 
${\mx}$ and the time variable $t$ are implicitly assumed, 
for instance
$ {\phi'}\partial_t{\phi}  =\int\! \dRM t \!\int \dRM^d{x}\, \phi'(t,\mx)\partial_t\phi(t,\mx)$.
 The term $\S_\text{mag}$ in the action~(\ref{eq:full_action1}) takes form
\begin{equation}  
\S_\text{mag}  = 
\frac{1}{2} \theta_i' D^\theta_{ij} \theta_j' + \theta_k'[ 
-\partial_t \theta_k -(v_i\partial_i)\theta_k + (\theta_i\partial_i)v_k + \nu_0 w_0 \partial^2 \theta_k],
\label{eq:mag_action} 
\end{equation}
where for convenience we have introduced new dimensionless parameter $w_0$ via $\kappa_0=\nu_0 w_0$, and
$D^\theta_{ij}$ denotes correlation function~\eqref{eq:noise1}.
On the other hand, advection of the tracer field corresponds to the field-theoretic action 
\vspace{-.4em}
\begin{align}
\S_2 & = \S_\text{vel}  + \S_\text{tracer}, 
\label{eq:full_action2} 
\end{align}
where $\S_\text{vel}$ is given by Eq.~\eqref{eq:vel_action} and $\S_\text{tracer}$ reads 
\vspace{-.4em}
\begin{equation}  
\S_\text{tracer}  = 
\frac{1}{2} \theta' D^\theta \theta' + \theta'[ 
-\partial_t \theta -(v_i\partial_i)\theta + \nu_0 w_0 \partial^2 \theta]
\label{eq:tracer_action} 
\end{equation}
with $D^\theta$ being the correlation function~\eqref{eq:noise2}.

In a field-theoretic formulation various stochastic
quantities (corresponding to Green functions in quantum field theory)
are calculable as functional integrals with a given weight functional $\exp\S$. 
Main benefits of such approach are transparency of a perturbation theory in Feynman graphs
and feasibility of the other powerful methods such as 
renormalization group and operator product expansion~\cite{Vasiliev,Tauber,Zinn}.
 
It is convenient to express the propagators of the theory in momentum-frequency representation 
\vspace{-1.0em}
\begin{align}
\langle v_iv_j' \rangle_{0} & =  {\langle v_j'v_i \rangle}^*_{0}= P_{ij}({\bm k})
\epsilon_{1}^{-1} + Q_{ij}({\bm k}) \epsilon_{3} R^{-1} ,
&\langle  v_j\phi' \rangle_{0}& = {\langle \phi'v_j \rangle}^*_{0}= 
-\frac{{\rm i} k_j}{R}, 
\label{eq:lines1}
\\
\langle v_iv_j \rangle_{0} & =   P_{ij}({\bm k}) \frac{d_1^{f}}{|\epsilon_{1}|^{2}}
+ Q_{ij}({\bm k}) d_2^{f} \left|\frac{\epsilon_{3}}{R}\right|^{2}, 
&\langle \phi v_j' \rangle_{0} & =  {\langle v_j' \phi \rangle}^*_{0}= -
\frac{{\rm i}c_{0}^{2} k_j}{R}, 
\label{eq:lines2}
\\
\langle  \phi\phi' \rangle_{0} & =  {\langle \phi'\phi \rangle}^*_{0}=
\frac{\epsilon_{2}}{R}, 
&\langle  \phi\phi \rangle_{0}& = \frac {c_{0}^{4} k^{2}d_2^{f}}{|R|^{2}},
\label{eq:lines3}
 \\
\langle  v_i\phi \rangle_{0} & =  {\langle  \phi v_i \rangle}^*_{0} =
\frac{ {\rm i} c_{0}^{2} d_2^{f}\epsilon_{3} k_i} {|R|^{2}} ,
\label{eq:lines4}
\end{align}
where the symbol ${z}^*$ denotes the complex conjugate of the expression $z$ and the following abbreviations have been used:
\vspace{-.8em}
\begin{align}
d^{f}_1  &=  g_{10}\nu_0^{3}\, k^{4-d-y}+g_{20}\nu_0^3, 
&d^{f}_2  &=  \alpha  g_{10}\nu_0^{3}\, k^{4-d-y}+g_{20}\nu_0^3, 
&\epsilon_{1}  & = -{\rm i}\omega +\nu_0 k^{2}, \nonumber\\
\epsilon_{2}&=-{\rm i}\omega+ u_0\nu_0 k^{2} , 
&\epsilon_{3}  & =  -{\rm i}\omega+ v_0\nu_0 k^{2} ,  
&R&=\epsilon_{2}\epsilon_{3}+c_{0}^{2}k^{2}. 
\label{R}
\end{align}

\begin{figure}[t]
\centerline{
\includegraphics[width=0.8\textwidth]{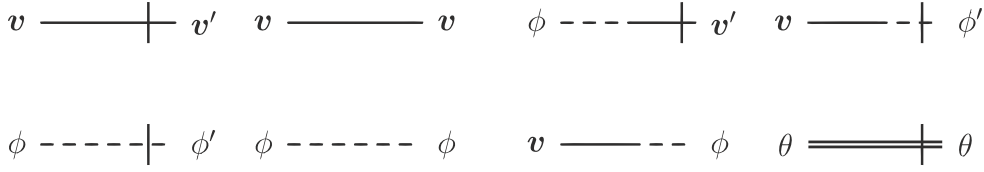}}
\caption{Graphical representation of all propagators of the models given by the quadratic part of the 
actions~(\ref{eq:full_action1}) and~(\ref{eq:full_action2}). }
\label{fig:prop}
\end{figure}

\begin{figure}[b]
\centerline{
\includegraphics[width=0.8\textwidth]{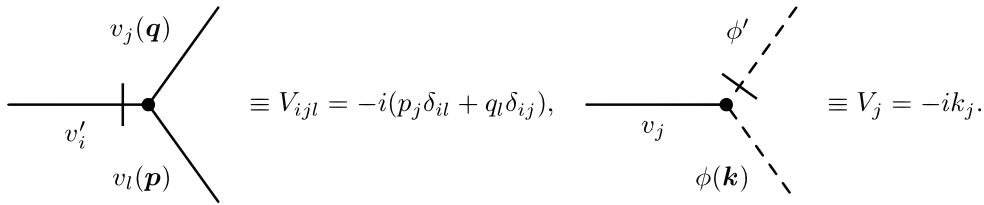}}
\caption{Graphical representation of all interaction vertices of the model related 
velocity non-linearities of the action~(\ref{eq:vel_action}).
For brevity we have retained only those momentum arguments that appear
in a resulting vertex factors.}
\label{fig:vert1}
\end{figure}

\begin{figure}[t]
\centerline{
\includegraphics[width=0.7\textwidth]{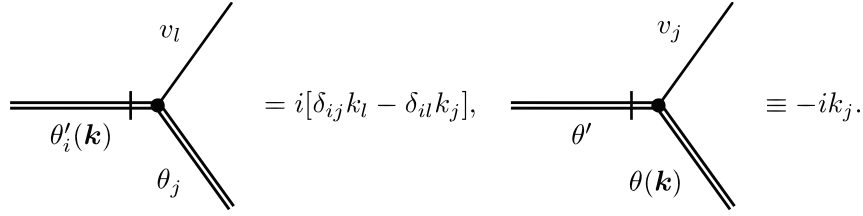}}
\caption{Graphical representation of all interaction vertices of the model given by the 
advection terms derived from action~(\ref{eq:mag_action})
and action~(\ref{eq:tracer_action}), respectively.}
\label{fig:vert2}
\end{figure}

\noindent
Graphical representation of propagators is depicted in Fig.~\ref{fig:prop}. Remaining 
propagators for magnetic admixture and tracer field take the following form, respectively
\begin{align}
 \langle \theta_i \theta_j' \rangle _0 & =  \langle \theta_j' \theta_i \rangle_0^* =
\frac{P_{ij}(\mk)} {-{\rm i}\omega +\kappa _0 k^2},
\label{eq:prop_tracer} \\
 \langle \theta \theta' \rangle _0 & =  \langle \theta' \theta \rangle_0^* =
\frac{1} {-{\rm i}\omega +\kappa _0 k^2}.
\label{eq:prop_vector}
\end{align}
There are two additional non-zero propagators $\langle \theta_i \theta_j \rangle _0$ and $\langle \theta \theta \rangle _0 $,
but in actual calculations they are in fact not needed~\cite{AK14,AK15}. Therefore, we do not quote them here.
All the remaining propagators are identically zero, i.e.,
$\langle  \phi'\phi' \rangle_{0}  =  \langle  v_i'\phi' \rangle_{0} =
\langle  v_i'v_j' \rangle_{0} = 
\langle  \theta_i'\theta_j' \rangle_{0} =
\langle  \theta' \theta' \rangle_{0} =  0$.
Self-explanatory graphical representations of non-linearities in studied models together with their vertex factors
are depicted in Fig.~\ref{fig:vert1} and Fig.~\ref{fig:vert2}.

Since the theory of critical phenomena corresponds to limit $k\to0$, in accordance with general theory all the terms in actional functional (and, as a consequence, in propagators) should have the same IR behavior. From the quantity $R$ in Eqs.~\eqref{R} it follows that $c_0^2k^2\simeq \epsilon_2\epsilon_3$. This means, that $c_0$ is IR significant parameter and, moreover, $c_0\simeq k$. 
Thus, $c_0\to0$ in IR asymptotic and considered model corresponds to large (at least not small as mentioned in~\cite{ANU97}) Mach numbers. The situation is typical for processes occuring in the solar corona but not in the atmosphere of the Earth. 
Note, that this requirement is not connected with calculation scheme which we use (MS scheme, see below) and which allows us to put $c_0=0$ in Feynman graphs just as a simplest way to performe calculations and give no restriction for the model. 
Here, we deal with basic feature of the theory; the situation is analogous to well-known $\varphi^4$ model, where parameter $\tau=T-T_c$ is IR significant and requirement $\tau\to0$ near the critical point is in accordance with both physical meaning and requirement of the model.
This is the reason why it is impossible to compare our results with previous works~\cite{VN96,ANS95} where corrections in small Mach numbers to the incompressible case were calculated.

%---------------------------------------------------------------------------------------%
\section{Renormalization group analysis 
\label{sec:RG}}

Theoretical models~(\ref{eq:full_action1}) and~(\ref{eq:full_action2}) are amenable to a
standard loop expansion using well-known Feynman diagrammatic rules~\cite{Vasiliev,Tauber}. However, as
it is often the case, ordinary perturbation theory is plagued by divergences and these must be properly taken
care of. The help comes from renormalization group method, which is considered here in dimensional regularization
within minimal subtraction (MS) scheme.

From a practical point of view, theory is renormalized once all
1-particle irreducible Green functions $\Gamma$ (further referred to as 1-irreducible functions)
are finite~\cite{Vasiliev,Tauber}. 
For dynamical models 
two independent scales have to be introduced: the time scale $T$ and the length scale $L$. 
Canonical dimensions of model parameters are found from the requirement that each term of the
action functionals~(\ref{eq:full_action1}) and~(\ref{eq:full_action2}) be dimensionless quantity with respect to 
both the momentum and 
frequency scales separately. We adopt  standard normalization conditions
\begin{equation}
d_k^k =-d_{ x}^k=1,\quad d_k^{\omega} =d_{x}^{\omega }=0,\quad
d_{\omega }^{\omega }=-d_t^{\omega }=1,\quad d_{\omega }^k=d_t^k=0.
\label{eq:def_dim}
\end{equation}
Then, the overall (total) canonical dimension of any quantity $F$ is
described by two numbers, the
frequency dimension $d_{F}^{\omega}$ and the momentum dimension $d_{F}^{k}$,
and given quantity $F$ therefore scales as 
$[F] \sim [T]^{-d_{F}^{\omega}} [L]^{-d_{F}^{k}}.$

Based on $d_F^k$ and $d_F^\omega$,  the total canonical dimension
$d_F=d_F^k+2d_F^\omega$ can be introduced, which in
the renormalization theory of dynamic models plays the
same role as the conventional (momentum) dimension does in
static problems~\cite{Vasiliev, Zinn}. 
Assuming quadratic dispersion law $\omega \sim k^{2}$ brought about a scaling relation between 
time and spatial scale which ensures that all the viscosity and diffusion coefficients in the model are dimensionless. 

The canonical dimensions for the velocity part of the model~(\ref{eq:vel_action}) are given in
Tab.~\ref{tab:vel}, whereas parameters of the magnetic and tracer part are given in Tab.~\ref{tab:mag}. From 
Tabs.~\ref{tab:vel} and \ref{tab:mag} it directly follows that the coupling constants $g_{10} \sim [L]^{-y}$
and $g_{20} \sim [L]^{-\varepsilon}$
become simultaneously dimensionless at $y=\varepsilon=0$ what corresponds to the logarithmic theory.
Since we use MS scheme the ultraviolet (UV) divergences in the
Green functions manifest
themselves as poles in $y$, $\varepsilon$ and their linear combinations.

\begin{table*}[t]
\centering
\caption{Canonical dimensions of the fields and parameters entering field-theoretic
action~(\ref{eq:vel_action}) for velocity fluctuations.}
\label{tab:vel}
\begin{tabular}{|c|c|c|c|c|c|c|c|c|c|}
    \hline
    $F$ & $ v_i'$ & $ v_i$ & $\phi'$ & $\phi$  &
     $\nu_0$, $\nu$ & $c_{0}$, $c$ &
    $g_{10}$ & $g_{20}$ & $u_{0}$, $v_{0}$, $u$, $v$, $g_1$, $g_2$, $\alpha$  \\
    \hline \hline
    $d_{F}^{k}$ & $d+1$ & $-1$ & $d+2$ & $-2$  
    &   $-2$ & $-1$ & $y$ & $4-d$ & 0 \\
    $d_{F}^{\omega}$ & $-1$ & 1 & $-2$ & 2 & 
     1 & 1 & 0 & 0 & 0\\
    $d_{F}$ & $d-1$ & 1 & $d-2$ & 2  & 0 & 1 & $y$ & $4-d$ & 0 \\
    \hline
  \end{tabular}
\end{table*}

\begin{table*}
\centering
\caption{Canonical dimensions of the fields and parameters entering actions
~(\ref{eq:mag_action}) and~(\ref{eq:tracer_action}) for advecting fields
 $\theta_i$ (magnetic field) and $\theta$ (tracer field), respectively.}
\label{tab:mag}
  \begin{tabular}{|c|c|c|c|c|c|c|c|c|c|c|}
  \hline
    $F$ & $ \theta_i', \theta' $ &  $\theta_i, \theta$ &
     $\kappa$, $\kappa_0$  & 
       $w_{0}$, $w$  \\
       \hline \hline
    $d_{F}^{k}$  & $d$  & 0
    &   $-2$  & 0 \\
    $d_{F}^{\omega}$ & $1/2$ & $-1/2$ &
     1  & 0\\
    $d_{F}$ & $d+1$ & $-1$ & 0  & 0 \\
    \hline
  \end{tabular}
\end{table*}

The total canonical dimension of any 1-irreducible 
function $\Gamma$ is given by the relation
\begin{equation}
\delta_{\Gamma} = d+2 - \sum_{\Phi} N_{\Phi} d_{\Phi},
\label{eq:index}
\end{equation}
where $N_{\Phi}$ is the number of the given type of field {entering the function}
$\Gamma$, $d_{\Phi}$ is the corresponding total canonical dimension of field $\Phi$, and
the summation runs
over all types of the fields $\Phi$ entering the 1-irreducible function $\Gamma$, see~\cite{Vasiliev}.  

Superficial UV divergences whose removal requires counterterms can be present only in
those functions $\Gamma$ 
for which
the formal index of divergence $\delta_{\Gamma}$ is a non-negative integer.
A dimensional analysis should be augmented by the several additional considerations.
They are summarized in the previous works~\cite{AK14,AK15,AGKL17_pre} and we do not repeat them here.
The crucial property of studied models is Galilean invariance, whose direct
consequence is that fields $v'_i,v_i,\theta_i$ and $\theta$ are not renormalized.
Thus, 
models under considerations with the actions~(\ref{eq:full_action1})
and~(\ref{eq:full_action2}) are
renormalizable and the only graphs that are divergent and needed to be considered are
two-point Green functions. 
For a velocity part~(\ref{eq:vel_action}), the following graphs should be analyzed:
\begin{align} 
  &  \raisebox{-1.ex}{ \includegraphics[width=3.2truecm]{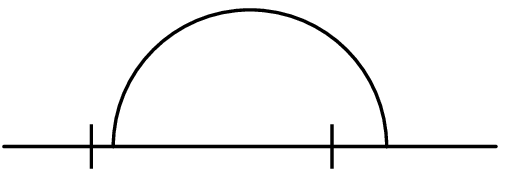}} 
     \raisebox{-1ex}{ \includegraphics[width=3.2truecm]{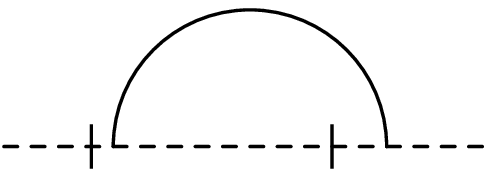}}
     \raisebox{-1ex}{ \includegraphics[width=2.75truecm]{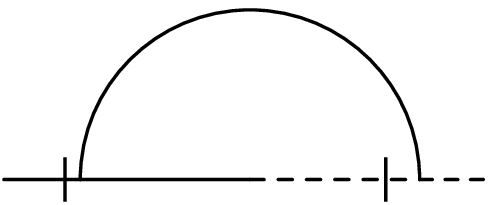}}
     \raisebox{-1ex}{ \includegraphics[width=3truecm]{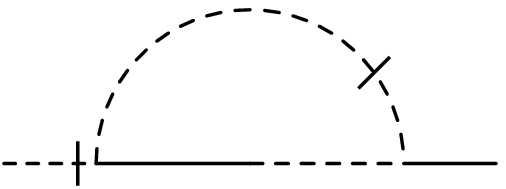}} \nonumber \\
  &  \raisebox{-1ex}{ \includegraphics[width=2.85truecm]{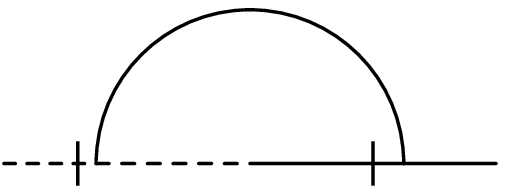}}      
     \raisebox{-1ex}{ \includegraphics[width=2.85truecm]{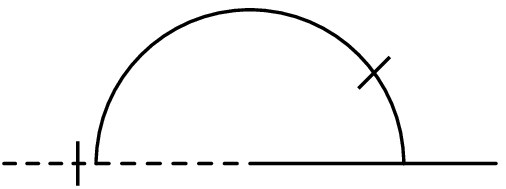}}
     \raisebox{-1.0ex}{ \includegraphics[width=3.1truecm]{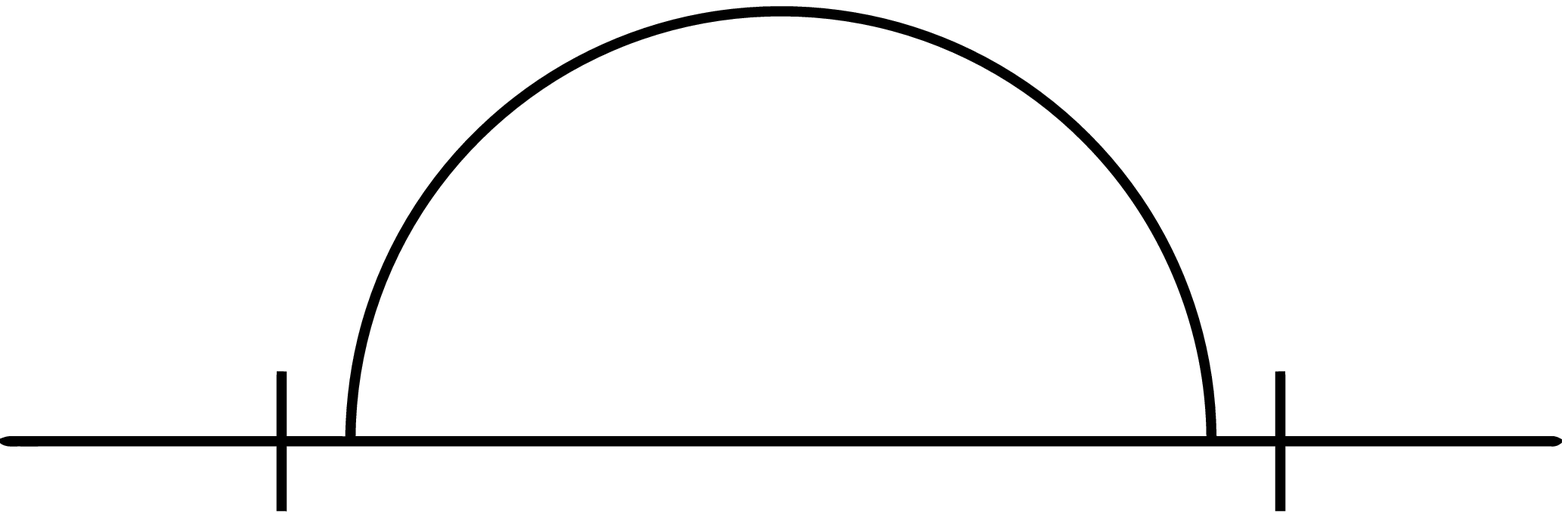}}.     
     \label{eq:vel_grafy}
\end{align}
For an advected part (vector or tracer field with appropriate changes in propagators and
 vertices) we have one additional graph:
\begin{align} 
\raisebox{-1.ex}{ \includegraphics[width=3.5truecm]{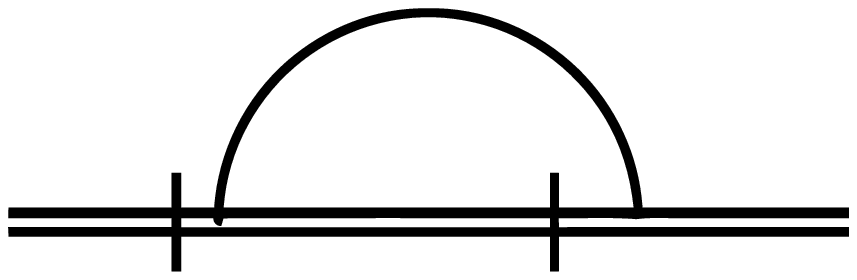}}. 
\label{eq:mag_grafy}
\end{align}
Remaining graphs are either UV finite or the Galilean invariance prohibits
their appearance. The calculation of the divergent parts of Feynman graphs proceeds in a straightforward
fashion and details can be found in~\cite{AK14,AK15,AGKL17_pre}. 

Although, the intermediate steps involved in calculation of the graph presented in~(\ref{eq:mag_grafy}) differs
for vector and tracer cases, the resulting divergent part of the graph is the same and reads
\begin{equation}   
   D =\frac{ \overline{S_d} } {2d}p^2
   \nu 
   \biggl\{
   \frac{1-d}{1+w}\biggl[ \frac{g_1}{y}\biggl(\frac{\mu}{m}\biggl)^y +
   \frac{g_2}{\eps}\biggl(\frac{\mu}{m}\biggl)^\eps 
   \biggl]
   {-
   \frac{u-w}{u(u+w)} 
   \biggl[ \frac{\alpha g_1}{y}\biggl(\frac{\mu}{m}\biggl)^y +
   \frac{g_2}{\eps}\biggl(\frac{\mu}{m}\biggl)^\eps 
   \biggl]   
  \biggl\} },
  \label{eq:mag_feynman1}
\end{equation}
where $\overline{S_d}=S_d/(2\pi)^d$ with $S_d = 2\pi^{d/2}/\Gamma(d/2)$ being the
surface area of the unit sphere in the $d$-dimensional space; $\Gamma(x)$ is
Euler's Gamma function. Parameter $\mu$ is a renormalization mass employed in minimal
subtraction scheme~\cite{Vasiliev,Zinn}. 
 For vector admixture expression~\eqref{eq:mag_feynman1} should be multiplied by
a projection operator $P_{ij}(\mpp)$ due to a vector nature of magnetic field $\theta_i$.

From the result~(\ref{eq:mag_feynman1}) we easily derive renormalization 
constant $Z_\kappa$ for the parameter $\kappa$ related to the advected fields [see Eq.~(\ref{eq:mag_action})]
and the anomalous dimension $\gamma_\kappa $:
\begin{align}
Z_\kappa & = 1 - \frac{g_1}{2dwy}\biggl[ 
\frac{d-1}{1+w} + \frac{\alpha(u-w)}{u(u+w)^2}
\biggl] - \frac{g_2}{2dw\eps} \biggl[ 
\frac{d-1}{1+w} + \frac{(u-w)}{u(u+w)^2}
\biggl],\\
\gamma_\kappa & = \frac{g_1}{2dw}\biggl[ 
\frac{d-1}{1+w} + \frac{\alpha(u-w)}{u(u+w)^2}
\biggl] + \frac{g_2}{2dw} \biggl[ 
\frac{d-1}{1+w} + \frac{(u-w)}{u(u+w)^2}
\biggl].
\end{align}

%---------------------------------------------------------------------------------------%
{\section{Scaling regimes} 
\label{sec:points}}

Underlying idea of  RG approach~\cite{Vasiliev,Tauber,Zinn} is embodied in a relation between 
the initial and renormalized action functionals 
$\S(\Phi,e_{0})= \S_{R}(Z_\Phi\Phi,e,\mu)$, where $e_{0}$
 denotes the complete set of bare parameters, $e$ is the set of their renormalized
counterparts, and $\Phi$ stands for a complete set of fields $\{\varphi\}$
and their response counterparts $\{\varphi'\}$. This relation can be converted
into a differential form
\begin{equation}
\biggl\{ {\cal D}_{RG} + N_{\varphi}\gamma_{\varphi} +
N_{\varphi'}\gamma_{\varphi'} \biggr\} \,G^{R}(e,\mu,\dots) = 0,
\label{eq:RG1}
\end{equation}
where $G =\langle \Phi\cdots\Phi\rangle$ is an arbitrary Green function of the theory;
$N_{\varphi}$ and $N_{\varphi'}$ are the numbers of entering fields $\varphi$ and $\varphi'$ in $G$,
the ellipsis commonly denotes  remaining arguments of $G$ (such as spatial and time variables),
${\cal D}_{RG}$ is the operation $\widetilde{\cal D}_{\mu}$
expressed in the renormalized variables, and
$\widetilde{\cal D}_{\mu}$
is the differential operation $\mu\partial_{\mu}$ at fixed bare parameters
$e_{0}$. For the present model it takes the form
\begin{equation}
  {\cal D}_{RG}= {\cal D}_{\mu} + 
  \sum \beta_g \partial_g
  - \gamma_{\nu}{\cal D}_{\nu}- \gamma_{c}{\cal D}_{c},
  \label{eq:RG2}
\end{equation}
where the sum runs over the set of all charges $\{ g_1,g_2,u,v,w\}$, $\nu$ and $c$ are viscosity and adiabatic speed of sound, respectively, and
differential operator ${\cal D}_{x} =x\partial_{x}$ has been introduced. 
The $\gamma$ and $\beta$-functions
are defined as
$\gamma_{F}= \widetilde{\cal D}_{\mu} \ln Z_F$
and $\beta_{g} = \widetilde{\cal D}_{\mu} g$.
The latter ones can be expressed in terms of anomalous dimensions:
\begin{equation}
  \beta_{g_1}  =  g_1\,(-y-\gamma_{g_1}), \,
  \beta_{g_2}  =  g_2\,(-\varepsilon-\gamma_{g_2}), \,
  \beta_{u}  =  -u\gamma_{u}, \,
  \beta_{v}  = -v\gamma_{v}, \,
  \beta_w  = w(\gamma_\nu - \gamma_\kappa).
  \label{eq:all_beta}
\end{equation}
The last expression follows from the introduced definition of the charge $w$ in Eq.~(\ref{eq:mag_action}).

Macroscopic scaling regimes are naturally identified with 
the IR attractive (``stable'') fixed points $g^*\equiv\{g_1^*,g_2^*,u^*,v^*,w^* \}$,
whose coordinates are found from the conditions~\cite{Vasiliev,Zinn}
\begin{align}
  &\beta_{g_1} (g^{*}) = \beta_{g_2} (g^{*})= \beta_{u}
  (g^{*}) = \beta_{v} (g^{*}) = \beta_w(g^*) = 0.
  \label{eq:gen_beta}
\end{align}
 This statement is a direct consequence of the invariant couplings behavior.
 Let us consider a set of invariant couplings $\overline{g}_i = \overline{g}_i(s,g)$
with the initial data $\overline{g}_i|_{s=1} = g_i$, where $s=k/\mu$. 
IR (macroscopic) asymptotic behavior is obtained in
the limit $s\rightarrow 0$. An evolution of invariant couplings satisfies flow equations
$\mathcal{D}_s \overline{g}_i = \beta_i(\overline{g}_j),$
and in a limit $s\to 0$ it can be approximated as 
$\overline{g}_i(s,g^*) \cong g^*+const\times s^{\omega_i}.$
A set $\{ \omega_i \}$ contains all eigenvalues of the matrix 
\begin{equation}
\Omega_{ij}=\partial\beta_{i}/\partial g_{j}|_{g=g_{*}}.
\label{eq:Omega}
\end{equation}
The existence of IR {attractive} solutions of the RG equations
corresponds to such fixed points for which the matrix~\eqref{eq:Omega} is positive definite.
These fixed points are then natural candidates for macroscopically observable scaling regimes. 
 
In the double expansion approach we have used, the character
of the IR behavior depends on the mutual relation between $y$ and $\varepsilon$.
In work~\cite{AGKL17_pre} the velocity part of the system~(\ref{eq:all_beta}), which don't include $\beta_w$, was 
thoroughly analyzed. The net result of the analysis is a prediction of three IR attractive fixed points.
The fixed point \fp{I} (the trivial or Gaussian point) is stable if $y$, $\varepsilon<0$ and has the coordinates 
\vspace{-.7em}
\begin{equation}
  g_1^* = 0, \quad g_2^* = 0, \quad \text{both}\ u^*\ \text{and}\ v^* \ \text{are arbitrary}.
  \label{fp1}
\end{equation}
The fixed point \fp{II}, which is stable if $\varepsilon>0$ and $y<3\varepsilon/2$, has the coordinates
\vspace{-.7em}
\begin{equation}
  g_1^* = 0, \quad g_2^* = \frac{8\eps}{3},\quad u^*=v^*=1.
  \label{eq:fp2}
\end{equation}
The fixed point \fp{III}, which is stable if $y>0$ and $y>3\varepsilon/2$, has the coordinates
\vspace{-.5em}
\begin{equation}
  g_1^* = \frac{16y(2y-3\eps)}{9[y(2+\alpha)-3\eps]}, 
  \quad g_2^* =\frac{16\alpha y^2}{9[y(2+\alpha)-3\eps]},\quad u^*=v^*=1.
  \label{eq:fp3}
\end{equation}
The crossover between the two nontrivial points~(\ref{eq:fp2}) and~(\ref{eq:fp3}) takes place across the
line $y=3\eps/2$, which is in accordance with results of~\cite{Ant04}.

Substituting obtained values of $u^*$ and $v^*$ together with $d=4$ we obtain for the charge $w$ the following beta function
\vspace{-1em}
\begin{equation}
   \beta_w = \frac{w-1}{16w(1+w)^2}   
   \biggl[ 
   g_1(6+2\alpha+9w+3w^2)+g_2(3w^2+9w+8)
   \biggl].   
   \label{eq:betaw}
\end{equation}
Note that this result agrees with previous works for the passive scalar case~\cite{AGKL17_pre}
and vector case~\cite{AHHJ03} as well.
The expression in the square brackets in Eq.~(\ref{eq:betaw}) is always positive
for physically permissible values $g_1 > 0,g_2>0,w>0$ and $\alpha>0$. Therefore, only
one nontrivial solution for the fixed point $w^*=1$ exists. It is
straightforward to prove that $\partial_w \beta_w >0$ for nontrivial fixed points \fp{II} and \fp{III}, which
ensures IR stability with respect to $w$ charge. Once scaling regimes are found, 
the critical dimensions for various quantities $F$ can be calculated via the relation	
\begin{equation}
  \Delta[F]=d^{k}_{F}+ \Delta_{\omega}d^{\omega}_{F} + \gamma_{F}^{*},
  \label{eq:Delta}
\end{equation}
where $d^{\omega}_{F}$ is the canonical frequency dimension, $d^{k}_{F}$ is the momentum dimension,
$\gamma_F^*$ is the anomalous dimension at the fixed point (FPII or FPIII), and
$\Delta_{\omega}=2-\gamma_\nu^*$ is the critical dimension of frequency.
Using Eq.~(\ref{eq:Delta}) the critical dimensions of advected fields are obtained for the fixed points \fp{II} and \fp{III}:
\begin{align}
  \Delta_{\theta_i} & = -1+\varepsilon/4, \quad \Delta_{\theta_i'}= d+1 -\varepsilon/4 \quad \text{for  \fp{II}},
  \label{eq:KriTet1} \\
  \Delta_{\theta_i} & = -1+y/6, \quad \Delta_{\theta_i'}= d+1 -y/6  \quad \text{for \fp{III}}.
  \label{eq:KriTet2}
\end{align}

%---------------------------------------------------------------------------------------%
{\section{ Composite Fields}
\label{sec:CF}}

From experimental point of view much more relevant than critical dimensions 
are quantities known as correlation and structure functions. In field-theoretic framework they are
usually represented by certain composite operators.
A local composite operator is a monomial or polynomial constructed from the primary fields
and their finite-order derivatives at a single space-time point. In the Green functions with such
objects, new UV divergences arise due to the coincidence of
the field arguments. Their removal requires an additional renormalization procedure~\cite{Vasiliev,Zinn}. 

It is not common to consider in one paper both tracer and vector admixtures. The reason is that
the expressions for tracer case are completely analogous to the vector ones in the part connected with composite fields $\theta$ and can
be easily obtained by considering operator $\partial_i\theta$ instead of $\theta_i$ in all formulas quoted below (note, that
propagators~\eqref{eq:prop_tracer}, \eqref{eq:prop_vector} and vertices depicted in Fig.~\ref{fig:vert2} still differs for tracer
and vector cases). This is a consequence of the fact that instead of density case for tracer admixture the operators $\theta^n$ are
UV finite and, therefore, we should consider operators $\partial_i\theta$ which contain derivatives. This is why both density and
tracer are scalar fields, but yield different expressions. Moreover, consideration of tracer field is much more closer to the vector case rather to the density one.

For brevity, hereinafter we write all expressions related to operators $\theta_i$ or $\partial_i\theta$
(for vector or tracer cases, respectively) below only for the vector case and use notation $\theta_i$ for vector field.
In the case of vector admixture we should focus on the irreducible tensor operators of the form
\begin{equation}
  F^{(n,l)}_{i_{1}\dots i_{l}} =
  \theta_{i_1}\cdots \theta_{i_l}\,
  ( \theta_j \theta_j)^{s} + \dots,
  \label{eq:Fnp}
\end{equation}
where $l$ is the number of the free vector indices (the rank of the tensor) and $n=l+2s$ is the total number 
of the fields $\theta_i$ entering the
operator. The ellipsis stands for the subtractions with the Kronecker's delta symbols that make the operator
irreducible (so that a contraction
with respect to any pair of the free tensor indices vanish). For instance,
\begin{equation}
  F^{(2,2)}_{ij} = \theta_i \theta_j -
  \frac{\delta_{ij}}{d}\, (\theta_k \theta_k).
  \label{eq:Irr}
\end{equation}

For a theoretical analysis, it is convenient to contract the tensors~(\ref{eq:Fnp}) with an arbitrary constant vector
{\mbox{\boldmath $\lambda$}}$=\{\lambda_{i}\}$.
The resulting scalar operator {then takes the form}
\begin{equation}
  F^{nl} = (\lambda_{i}w_{i})^{l} (w_{i}w_{i})^{s} + \dots,
  \quad w_{i} \equiv \theta_i,
  \label{eq:FnpSk}
\end{equation}
where the subtractions, denoted by the ellipsis, necessarily contain
factors of $\lambda^{2}=\lambda_{i}\lambda_{i}$. 

In order to calculate the critical dimension of an operator, we have to renormalize it. The 
operators~(\ref{eq:Fnp}) can be treated as
multiplicatively renormalizable, $F^{nl} = Z_{nl} F^{nl}_{R}$, with certain renormalization
constants $Z_{(n,l)}$~\cite{AK14,AK15}. 
The counterterm to $F^{nl}$ must have the same rank as the operator itself. It means that the terms 
containing $\lambda^{2}$  should be
excluded since the contracted fields $w_{i}w_{i}$, which naturally appear in such terms, reduce the number of free indices.
The renormalization constants $Z_{nl}$ are determined by the finiteness of the 1-irreducible Green 
function $\Gamma_{nl}(x; \theta)$, where $x=(t,\mx)$ is a functional argument of the operator $\theta$.
In  the one-loop approximation we have a diagrammatic equation
\begin{equation}
  \Gamma_{nl}(x; \theta) = F^{nl} + \frac{1}{2}   
   \raisebox{-4.ex}{ \includegraphics[width=1.2truecm]{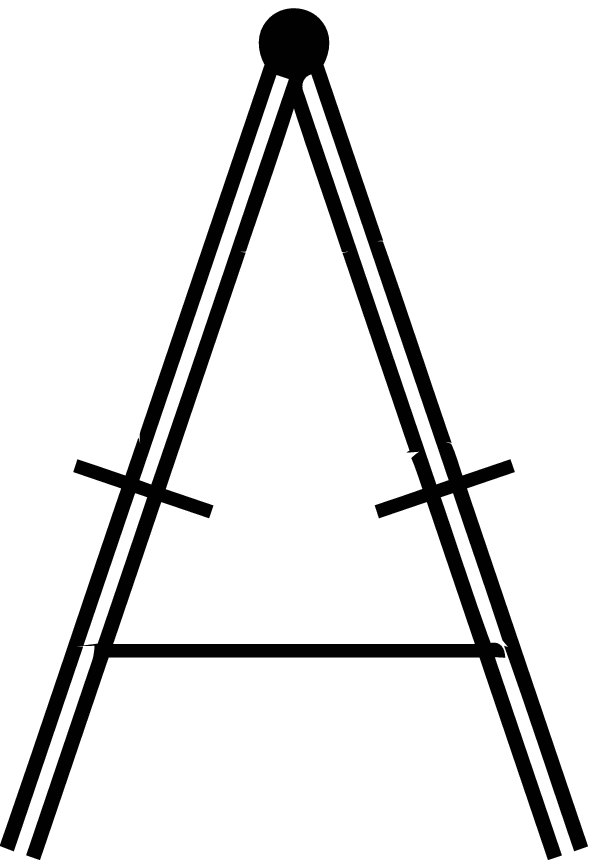}},
  \label{eq:dom}
\end{equation}
where numerical factor $1/2$ is a symmetry factor of the graph.
 The thick dot in the top of the graph represents the operator vertex, which up to irrelevant terms can be presented in the form 
\begin{equation}
  V(x;x_{1},x_{2})=  \frac{\partial^{2}F^{nl}}{\partial w_{i}\partial w_{j}}
  \, \delta(x-x_{1})\, \delta(x-x_{2}).
  \label{eq:Vae10}
\end{equation}
The differentiation yields
\begin{align}
  {\partial^{2}F^{nl}}/{\partial w_{i}\partial w_{j}} & = 
  2s (w^{2})^{s-2} (\lambda w)^{l} \left[\delta_{ij} w^{2} +2(s-1)w_{i}w_{j}
  \right] + l(l-1) (w^{2})^{s} \nonumber \\
  &\times (\lambda w)^{l-2} \lambda_{i} \lambda_{j}
   +  2ls (w^{2})^{s-1}(\lambda w)^{l-1} (w_{i}\lambda_{j}+ w_{j}\lambda_{i}),
  \label{eq:Vae11}
\end{align}
where $w^{2}=w_{k}w_{k}$, $(\lambda w)=\lambda_{k}w_{k}$
and substitution $w_i \rightarrow \theta_i$ is assumed.
Two more factors $w_{p}w_{r}$ are attached to the bottom of the graph due to the
derivatives stemming from the vertices $\theta'_i(v_k\partial_k)\theta_i$. 

The UV divergence is logarithmic and one can set all the external frequencies and
momenta equal to zero. Since propagators~\eqref{eq:prop_tracer}, \eqref{eq:prop_vector} and vertices depicted 
in Fig.~\ref{fig:vert2} are different for vector and tracer cases, the core of the graph also differs for these two cases. For tracer field it reads
\begin{equation}
  \widehat{D}_{ij}^{pr}=\int\frac{\text{d}\omega}{2\pi}
  \int\displaylimits_{k>m}\frac{{\dRM^d}{k}}{(2\pi)^{d}}\,
  k_{i}k_{j}\, D^v_{pr} (\omega, {\mk})\,
  \frac{1}{\omega^{2}+w^{2}\nu^2 k^{4}}.
  \label{eq:triad}
\end{equation}
Here, the first factor comes from the derivatives in~(\ref{eq:Vae11}),  $D^v_{pr}$ is the velocity
correlation function [see Eq.~(\ref{eq:correl2})], and the last factor comes from the two propagators $\langle\theta'_i\theta_j\rangle_{0}$.
 The indices $i$ and $j$ should be later contracted with expression~(\ref{eq:Vae11}), the indices $p$ and $r$ with 
external fields $w_p$ and $w_r$ denoted by ``legs'' of the graph. 

For vector field the core of the graph takes the form
\begin{equation}
 \widetilde{D}_{ij}^{xy}=\int\frac{\text{d}\omega}{2\pi}
  \int\displaylimits_{k>m}\frac{{\dRM^d}{k}}{(2\pi)^{d}}\,
  P_{ai}(\bm k)P_{bj}(\bm k)\, D^v_{pr} (\omega, {\mk})\, 
  \frac{1}{\omega^{2}+w^{2}\nu^2 k^{4}}V_{apx}(\bm k)V_{bry}(\bm k),
  \label{eq:triad_vec}
\end{equation}
where $P_{ai}(\bm k)$ and $P_{bj}(\bm k)$ follows from propagators~\eqref{eq:prop_vector}, $V_{apx}(\bm k)$ and $V_{bry}(\bm k)$ 
are two vector vertices (see Fig.~\ref{fig:vert2}), and $D^v_{pr}$ is the velocity correlation function. For brevity we do not draw 
here picture for the graph with explicitly written vector indices, but they may be easily restored from the above written expression. 

 After the integration, combining all the
factors, contracting the tensor indices and expressing the result in terms of $n=l+2s$ and $l$, one obtains
\begin{align}   
  \raisebox{-4.ex}{ \includegraphics[width=1.2truecm]{comp.eps}}
  & = -\frac{F_{nl}}{2wd(d+2)}\biggl[  
  \frac{ Q_1}{(1+w)} \biggl( \frac{g_1}{y} + \frac{g_2}{\eps} \biggl)
  +
  \frac{ Q_2}{u(u+w)} \biggl( \frac{\alpha g_1}{y} + \frac{g_2}{\eps} \biggl)
  \biggl],
\end{align}
where
\begin{align}   
  Q_1 &=-n(n+d)(d-1)+l(d+1)(d+l-2), \nonumber \\
  Q_2 & =-n(3n+d-4)+l(d+l-2)
  \label{Q-tr}
\end{align}
for tracer case and 
\begin{align}   
  Q_1 &=-n(n+d)(d-1)+l(d+1)(d+l-2), \nonumber \\
  Q_2 & =-n(n+nd-d)(d-1)+l(d+l-2)
  \label{Q-vec}
\end{align}  
 for vector case\footnote{ Note that in previous study~\cite{AK15} there are misprints in expressions~(5.18)
for quantities $Q_1$ and $Q_2$. Right expressions are Eq.~\eqref{Q-vec} written above and Eq.~(5.7) in~\cite{AHHJ03}.}. 
Finally, expression~\eqref{eq:dom} reads
\begin{equation}
  \Gamma_{nl}(x;\theta) = F^{nl}(x)\, \biggl\{ 1 - \frac{1}{4wd(d+2)}\,
  \biggl[
  \frac{Q_{1}}{(1+w)} \biggl( \frac{g_1 }{y} + \frac{g_2 }{\eps}\biggl)
   + 
  \frac{Q_{2}}{u(u+w)}\biggl( \frac{\alpha g_1}{y} +  \frac{ g_2 }{\eps}\biggl)
  \biggl] \biggl\}.
  \label{eq:QQ}
\end{equation}
The renormalization constants $Z_{nl}$ calculated in the MS scheme thus take the form
\begin{equation}
  Z_{nl} = 1 - \frac{1}{4dw(d+2)}\,
  \left[
  \frac{Q_{1}}{1+w} \left(\frac{ g_{1} }{y}+\frac{ g_{2} }{\eps}\right) + 
  \frac{Q_{2}}{u(u+w)}\left( \frac{\alpha g_{1} }{y}+\frac{ g_{2} }{\eps}\right)
  \right]
  \label{eq:ZZ}
\end{equation}
and for the corresponding anomalous dimensions we  get
\begin{equation}
  \gamma_{nl} = \frac{1}{4dw(d+2)}\, \left\{
  \frac{Q_{1}}{1+w} (g_{1}+g_{2}) +  \frac{Q_{2}}{u(u+w)} (\alpha g_{1}+g_{2})
  \right\}.
  \label{eq:GG}
\end{equation}

Now in order to evaluate the critical dimensions of the operators $F_{nl}$ one needs to substitute the coordinates of the fixed 
points into the  
expression~(\ref{eq:GG}) and then use the relation~(\ref{eq:Delta}). For the fixed point FPII the critical dimension is
\begin{equation}
  \Delta_{nl} = \frac{n}{4}\varepsilon + \frac{Q_1+Q_2}{72}\varepsilon;
  \label{eq:Dnl1}
\end{equation}
for the fixed point FPIII it is
\begin{equation}
  \Delta_{nl} = \frac{n}{6}y + \frac{y}{12}\frac{Q_1 (\alpha y + 2y - 3\varepsilon) + 3\alpha Q_2 (y-\varepsilon)}{9[y(2+\alpha)-3\varepsilon]}.
  \label{eq:Dnl2}
\end{equation}
Both expressions~\eqref{eq:Dnl1} and~\eqref{eq:Dnl2} might be affected by higher order corrections in $y$ and 
$\varepsilon$.  Inspection of expressions~\eqref{eq:Dnl1} and~\eqref{eq:Dnl2} reveals 
that increasing value of $n$ leads to a  infinite set of  operators with negative critical dimensions. Their 
spectra are unbounded from below, see Appendix~A in~\cite{AK15}. 

The latter result for \fp{III} is in accordance with previously known results~\cite{AK14,AK15} for the analysis near three-dimensional space $d=3$:
\begin{align}
\Delta_{nl}&=\frac{ny}{6}+\frac{y(Q_1+\alpha Q_2)}{6(d-1)(d+2)},
\label{15}
\end{align}
where $Q_1$ and $Q_2$ coincide with those given in Eqs.~\eqref{Q-tr} and~\eqref{Q-vec} for tracer and vector cases, respectively. Expression~\eqref{15} at $d=4$ reads
\begin{align}
\label{KrOp2-1}
\big.\Delta_{nl}\biggl|_{d=4}&=\frac{ny}{6}+\frac{y(Q_1+\alpha Q_2)}{108}.
\end{align}
Expanding the  expression~\eqref{eq:Dnl2} in $y$ at fixed value  $\eps=1$ (which corresponds to $d=3$) yields
\begin{align}
\label{KrOp2-2}
\Delta_{nl}&=\frac{ny}{6}+\frac{y(Q_1+\alpha Q_2)}{108}+\mathcal{O}(y^2).
\end{align}
From the expressions~\eqref{KrOp2-1} and~\eqref{KrOp2-2} it follows that the expression~\eqref{eq:Dnl2}, obtained as a result of the double $y$ and
$\varepsilon$ expansion near $d=4$, may be considered as a certain partial infinite resummation of the ordinary $y$ expansion. This
resummation significantly improves the situation at large $\alpha$: now we do not have the pathology when the critical dimensions
$\Delta_{nl}$ are linear in $\alpha$ and, therefore, grow with $\alpha$ without a bound.

%---------------------------------------------------------------------------------------%
{\section{Operator Product Expansion} 
\label{sec:OPE}}

Our main interest are pair correlation functions constructed from the composite operators, whose
unrenormalized counterparts have been defined in Eq.~(\ref{eq:Fnp}).
For Galilean invariant equal-time functions we can write
the representation
\begin{equation}
  \langle F^{mi}(t,\mx) F^{nj} (t,\mx') \rangle \simeq \mu^{d_F} \nu^{d^\omega_F} (\mu r)^{-\Delta_{mi}-
  \Delta_{nj}} \zeta_{mi;nj}(mr, c(r)),  
  \label{eq:struc}
\end{equation}
where $r=|\mx-\mx'|$ and $c(r)$ is effective speed of sound. Its limiting behavior is
\begin{equation}
  c(r) = c \frac{(\mu r)^{\Delta_c}  }{\mu\nu} \rightarrow
         \begin{cases}
           c(0) \quad \text{{for} non-local regime FPIII}; \\
           c(\infty) \quad \text{{for} local regime FPII},
         \end{cases}
\end{equation}
see~\cite{AGKL17_pre}.

Expression~(\ref{eq:struc}) is valid  in the
asymptotic limit $\mu r\gg 1$. Further, the inertial-convective range corresponds to the additional
restriction $mr\ll 1$. The behavior of the
functions $\zeta$ at $mr\to0$ can be studied by means of the OPE technique~\cite{Vasiliev,Zinn}.
The basic idea of this method is to represent a product of two operators
at two close points $\mx$ and $\mx'$ with a condition $\mx-\mx'\rightarrow 0$ in the form
\begin{equation}
  F^{mi}(t,\mx)F^{nj}(t,\mx') \simeq \sum_{F} C_{F}(mr)\,
  F \biggl(t,\frac{\mx + \mx'}{2} \biggl),
  \label{eq:OPE2}
\end{equation}
where functions $C_F$ are regular in their arguments and a given sum runs over
all permissible local composite operators $F$ allowed by RG and symmetry considerations.
Taken into account~(\ref{eq:struc}) and~(\ref{eq:OPE2}) in the limit $mr \rightarrow 0$  we arrive
at the relation
	\begin{equation}
	   \zeta(mr) \simeq \sum_F  A_F(mr)
	   (mr)^{\Delta_F}.
	   \label{ansc}
	\end{equation}
Singularities for $mr\to0$  (and thus the anomalous scaling) result from the contributions in~\eqref{ansc} of the 
operators with negative critical dimensions, see Eqs.~\eqref{eq:Dnl1} and~\eqref{eq:Dnl2}. 
	
Considering OPE for the correlation functions $\langle F^{(p,0)} F^{(k,0)} \rangle$ 
constructed from scalar operators of the type~(\ref{eq:Fnp}), one
can observe that the leading contribution to the expansion is determined by the
operator $F^{(n,0)}$ with $n=p+k$ from 
the same family.
Therefore, in the inertial range these correlation functions acquire the form
\begin{equation}
  \langle F^{(p,0)}(t,{\mx}) F^{(k,0)}(t,{\mx'}) \rangle \sim
  r^{-\Delta_{p0}-\Delta_{k0}+\Delta_{n0}}.
  \label{eq:MF}
\end{equation}
The inequality $\Delta_{n0}<\Delta_{p0}+\Delta_{k0}$, which follows from both explicit one-loop
expressions~(\ref{eq:Dnl1}) and~(\ref{eq:Dnl2}), indicates,
that the operators $F^{(n,0)}$ demonstrate a multifractal behavior for both regimes FPII and FPIII; see~\cite{DL,DL-2}.

A direct substitution of $d=4$ leads to the following prediction for a critical dimensions
\begin{equation}
   \Delta_{nl} = n\Delta_\theta + \gamma^*_{nl} = 
   \begin{cases}
      -n + \frac{n\eps}{4} +\frac{(Q_1+Q_2)\eps}{4} \qquad\mbox{ }\hskip1.6cm \text{for FPII}, \\
      -n + \frac{n y}{6} + \frac{Q_1 y}{108} + \frac{Q_2 \alpha y(y-\eps)}{36[y(2+\alpha)-3\eps]}
      \qquad \text{for FPIII},
   \end{cases}
   \label{eq:OPE_res}
\end{equation}
where we have
\begin{equation}
  Q_1|_{d=4} = -3n(n+4)+ 5l(l+2) ,\quad 
  Q_2|_{d=4} = -3n^2+l(l+2)
\label{eq:OPE_res2-tr}
\end{equation}
for tracer case and 
{\begin{equation}
  Q_1|_{d=4} = -3n(n+4)+ 5l(l+2) ,\quad 
  Q_2|_{d=4} = -3n(5n-4)+l(l+2)
\label{eq:OPE_res2-vec}
\end{equation}
 for vector case. 
 From Eq.~\eqref{eq:OPE_res} and Eqs.~\eqref{eq:OPE_res2-tr}, \eqref{eq:OPE_res2-vec} it follows that for fixed $n$ a kind of an hierarchy present with respect to
the index $l$, which measures the ``degree of anisotropy:''
\begin{equation}
\frac{\partial \Delta_{nl} }{\partial l} >0.
\end{equation}
In other words, the higher $l$ the less important contribution. The most relevant
is given by the isotropic shell with $l=0$. This is in accordance with
previous studies~\cite{AK14,AK15,AHHJ03} and hypothesis about restoration of isotropy and symmetries of turbulent motion in 
the statistical sence in the depth of inertial interval~\cite{AGKM18}.

%-------------------------------------------------------------------------------------------%
{\section{Conclusion} 
\label{sec:end}}

In the present paper the advection problem of the vector and tracer field by the Navier-Stokes velocity
ensemble have been examined. The fluid was assumed to be
compressible and the space dimension was close to $d=4$. This specific case allows us to take into consideration one more 
divergent function, namely $\left\langle v'v'\right\rangle_{\text{1-ir}}$, and construct the double expansion in $y$ and $\varepsilon=4-d$.
This leads to richer results in comparison with naive consideration of the system near physical dimension $d=3$.

 Using renormalization group technique two nontrivial IR stable fixed points were identified and, therefore, the critical behavior in the inertial 
range demonstrates two different nontrivial regimes depending on the 
relation between the exponents $y$ and $\varepsilon$. The expressions for the critical exponents of the advected 
fields $\theta$ were obtained in the leading one-loop approximation. 

In order to find the anomalous exponents of the structure functions, the composite fields~(\ref{eq:Fnp}) were 
renormalized. The critical dimensions of them were evaluated. 
Moreover, operator product expansion allowed us to derive the 
explicit expressions for the critical dimensions of the structure functions. 

The existence of the anomalous scaling (singular 
power-like dependence on the integral scale $L$) in the inertial-convective range was established for both possible
scaling regimes.
 From the leading order (one-loop) 
calculations it follows that the main contribution into the OPE is given by the isotropic term 
corresponding to $l=0$, where $l$ is the rank of the tensor and serves as a degree of the anisotropy; all other terms with $ l \geq 2 $ provide 
only corrections.
These facts give a quantitative illustration of the general concept that the symmetries of the Navier-Stokes 
equation, broken spontaneously and by initial or boundary conditions, are restored in the statistical 
sense for the fully developed turbulence~\cite{Frisch,Monin,Davidson}. 
Another very interesting result is that correlations of the advected fields exhibit multifractal behavior.

 The results of this study are especially significant at large values of $\alpha$ (purely potential random force).
In contrast to analysis near $d=3$, in the present case the anomalous dimensions of the composite operators~\eqref{eq:Dnl1} and~\eqref{eq:Dnl2} do not 
grow with $\alpha$ without a bound. This is a consequence of eliminating the poles in $\varepsilon$ near $d=4$, which leads 
to a significant improvement of calculated expressions for critical dimensions near physical value \mbox{$d=3$}.
Expression~\eqref{eq:Dnl2} obtained in this study may be considered as an example of infinite resummation of ordinary $y$ expansion. 
It works well at large $\alpha$ being not expanded in $y$, and the first 
term of this expansion coincides with the answer presented earlier in~\cite{AK15}; see expressions~\eqref{KrOp2-1}~-- \eqref{KrOp2-2}.

Regarding future tasks, it would be interesting to go beyond the one-loop approximation and to 
analyze the behavior more accurately.
Another very important issue to be further 
investigated is to have a closer look at the both scalar and vector active fields, 
i.e., to consider a back influence of the advected fields to the turbulent environment flow.

%-------------------------------------------------------------------------------------------%
 {\section*{Acknowledgments} 

T.~L. gratefully acknowledge financial support provided by  VEGA grant No. $1/0345/17$ 
of the Ministry of Education, Science, Research and Sport of the Slovak Republic and
the grant of the Slovak Research and Development Agency under the contract No.~APVV-16-0186.
 The work was supported by the Russian Foundation for Basic Research within the
Project No.~18-32-00238 (all results concerning MHD and vector admixture) and by the Foundation for the Advancement of Theoretical Physics and Mathematics ``BASIS.'' 
N.M.G. acknowledges the support from the Saint Petersburg Committee of Science and High School.

%-------------------------------------------------------------------------------------------%

\end{document}